# Momentum-Dark Intervalley Exciton in Monolayer Tungsten Diselenide Brightened *via* Chiral Phonon

Zhipeng Li[1,2#], Tianmeng Wang[2#], Chenhao Jin[3#], Zhengguang Lu[4,5], Zhen Lian[2], Yuze Meng[2,6], Mark Blei[8], Mengnan Gao[11], Takashi Taniguchi[7], Kenji Watanabe[7], Tianhui Ren[1*], Ting Cao[9,10], Sefaattin Tongay[8], Dmitry Smirnov[4], Lifa Zhang[11*], Su-Fei Shi[2,12*]

1. School of Chemistry and Chemical Engineering, Key Laboratory for Thin Film and Micro Fabrication of the Ministry of Education, Shanghai Jiao Tong University, Shanghai, 200240, China.
2. Department of Chemical and Biological Engineering, Rensselaer Polytechnic Institute, Troy, NY 12180.
3. Kavli Institute at Cornell for Nanoscale Science, Ithaca, NY, 14853.
4. National High Magnetic Field Lab, Tallahassee, FL, 32310.
5. Department of Physics, Florida State University, Tallahassee, Florida 32306.
6. College of Physics, Nanjing University, Nanjing, 210093, P. R. China.
7. National Institute for Materials Science, 1-1 Namiki, Tsukuba 305-0044, Japan.
8. School for Engineering of Matter, Transport and Energy, Arizona State University, Tempe, AZ 85287.
9. Geballe Laboratory for Advanced Materials, Stanford University, Stanford, CA 94305.
10. Department of Materials Science and Engineering, University of Washington, Seattle, WA 98195.
11. Center for Quantum Transport and Thermal Energy Science, School of Physics and Technology, Nanjing Normal University, Nanjing, 210023, China.
12. Department of Electrical, Computer & Systems Engineering, Rensselaer Polytechnic Institute, Troy, NY 12180.

## ABSTRACT

Inversion symmetry breaking and three-fold rotation symmetry grant the valley degree of freedom to the robust exciton in monolayer transition metal dichalcogenides (TMDCs), which can be exploited for valleytronics applications. However, the short lifetime of the exciton significantly constrains the possible applications. In contrast, dark exciton could be long-lived but does not necessarily possess the valley degree of freedom. In this work, we report the identification of the momentum-dark, intervalley exciton in monolayer WSe$_2$ through low-temperature magneto-photoluminescence (PL) spectra. Interestingly, the intervalley exciton is brightened through the emission of a chiral phonon at the corners of the Brillouin zone (K point), and the pseudoangular momentum (PAM) of the phonon is transferred to the emitted photon to preserve the valley information. The chiral phonon energy is determined to be ~ 23 meV, based on the experimentally extracted exchange interaction (~ 7 meV), in excellent agreement with the theoretical expectation of 24.6 meV. The long-lived intervalley exciton with valley degree of freedom adds an exciting quasiparticle for valleytronics, and the coupling between the chiral phonon and intervalley exciton furnishes a venue for valley spin manipulation.

In monolayer transition metal dichalcogenides (TMDCs), the inversion symmetry breaking and three-fold rotation symmetry endow the intravalley exciton valley degree of freedom, which can be accessed through circularly polarized light[1–9] and ushers in the exciting field of valleytronics.[3,4,16,5,6,10–15] However, the exciton in TMDCs is typically short-lived, with the lifetime of a few to tens of picoseconds,[17–19] limiting the potential applications. The spin-forbidden dark exciton in tungsten-based TMDCs[20–22] possesses a much longer lifetime and caught significant attention for the last few years. However, the out-of-plane radiation of the spin-forbidden dark exciton is not protected by the same valley physics of the in-plane dipole radiation of the bright exciton that is constrained by the three fold rotation symmetry and inversion symmetry breaking, and the valley degree of freedom cannot be easily maintained. Intervalley exciton, with the hole residing in the K valley and electron in the K' valley (Fig. 1), due to the restricted intervalley scattering of the hole, could potentially solve this problem. Intervalley exciton can naturally occur when excitons are selectively generated by optical pumping in the K valley (Fig. 1), as the electron can be relatively easy to be scattered to the K' valley, and intervalley exciton thus could be both long-lived due to the momentum mismatch and valley polarized due to the constrained intervalley scattering of the hole[23] because of the large valence band splitting induced by the spin-orbit coupling.

In this work, we identify the intervalley exciton in high-quality monolayer $WSe_2$ device through the gate and magnetic field dependent photoluminescence (PL) spectra at low temperature. We also unveil the intervalley exciton coupling with a specialized phonon, the chiral phonon of the $WSe_2$. Chiral phonons in monolayer TMDCs, also stemming from three-fold rotation symmetry, were theoretically predicted at Brillouin-zone corners.[24] The large momentum of the chiral phonon can resolve the dilemma of the momentum mismatch that prohibits the recombination of intervalley citons. In addition, the pseudoangular momentum (PAM) of the chiral phonon potentially can be transferred to and determine the helicity of the emitted photon through phonon-exciton interaction.

The intervalley exciton and chiral phonon coupling also allows us to directly study the chiral phonon. Chiral phonons play a critical role in intervalley scattering[25] and can be utilized for phononic chirality, phonon-driven topological states, and dissipationless information processing. However, direct probing of the chiral phonon at the Brillouin-zone corners (K and K' valleys) is restrained by the large momentum mismatch between the phonon mode and the photon, which prevents us from using polarization-resolved Raman spectroscopy to investigate the energy and chirality of the phonon mode directly. Very recently, the existence of such chiral phonon modes was verified by transient infrared spectroscopy.[25] However, the experimental complexity for chiral phonon creation and detection gives rise to uncertainty in determining the chiral phonon energy. In contrast, the energy conservation of the intervalley exciton recombination allows us to determine the phonon energy accurately by analyzing the emitted photon energy through PL spectra, and we experimentally extracted the phonon energy to be ~ 23 meV, in excellent agreement with the prediction from first principle calculations (24.6 meV, Supplementary Table S3).[25] The angular momentum conservation of the process also allows us to determine the phonon PAM to be 1 (inset of Fig. 1) by analyzing the emitted photon chirality in the helicity-resolved PL spectra.

Through the valley-resolved PL spectra, we found that the emitted photon retains the valley information due to the PAM transferred from the chiral phonon. And the time-resolved PL (TRPL) spectra show that the lifetime of the momentum-dark intervalley exciton is ~ 200 ps, comparable to that of the spin-forbidden dark exciton (~ 250 ps) but much longer than that of the bright exciton (~ 5 ps).[26] The valley-polarized, long-lived intervalley exciton thus adds to the excitonic quasiparticles for the valleytronics, and the coupling between the chiral phonon and the intervalley exciton can be utilized as a tunable knob for manipulating valley degree of freedom in TMDCs.

**RESULTS**

We fabricated a BN encapsulated monolayer WSe$_2$ device using a dry pickup method described in our previous studies.[26,27] The device structure is shown schematically in Fig. 2a, in which both few-layer graphene flakes were employed to work as the contact electrode to the WSe$_2$ and the semi-transparent top-gate electrode, respectively, with the top BN flake working as the gate dielectric. The top-gate voltage allows efficient control of carriers in WSe$_2$. The optical microscope image for a typical device is shown in Fig. 2b. The PL spectra for two different devices at 4.2 K are shown in Fig. 2(c,d), in which most of the features have been identified in the previous work, including the bright exciton ($X_0$), the biexciton (XX) and charge biexciton (XX$^-$),[27–34] the two negative trions ($X_1^-$ and $X_2^-$),[4,35–38] the dark exciton ($X_D$),[19–22,39,40] and the dark exciton phonon replica ($X_D^R$).[26] However, the origin of the PL peak at 1.674 eV in Fig. 2c and 1.681 eV in Fig. 2d remains unknown. Despite the variation of the PL peak position in the two different devices, the energy difference between this emerging peak and that of the dark exciton remains constant, 16 meV for both device 1 (Fig. 2c) and device 2 (Fig. 2d). We thus believe that the emerging PL peak in the two different devices shares the same origin and we label it as $X_i$. This hypothesis is also confirmed by the results from two other devices (see Fig. S1 in SI).

To explore the origin of the PL peak $X_i$, we performed PL spectra as a function of the top gate voltage, which effectively controls the density and type of the charge carriers in the monolayer WSe$_2$. As shown in Fig. 3a, the intensity of the emerging PL peak as a function of the gate voltage closely follows the gate dependence of the dark exciton peak $X_D$. We found the PL intensity of the peak $X_i$ is maximized between gate voltages -1.0 V to 0.35 V, when WSe$_2$ is charge neutral, which suggests that the $X_i$ stems from the photon emission of a charge-neutral quasiparticle. The gate dependence rules out the possibility of the Q-K exciton[41], whose PL will have different gate dependence from that of the spin-forbidden dark exciton ($X_D$). It is worth noting that at the n doping (gate voltage > 0.35 V) and p doping (gate voltage < -1.0 V) sides, PL peaks of negative dark trion ($X_D^-$) and positive dark trion ($X_D^+$) emerge (Fig. 3a).[22,42,43]

To further illustrate the nature of $X_i$, we performed PL spectra measurements as a function of the out-of-plane magnetic field. The PL spectra measurements were performed in a valley-resolved configuration in which we excited the monolayer WSe$_2$ with either right circularly polarized light or left circularly polarized light ($\sigma^-$), and we detected PL with the same polarization, *i.e.*, ($\sigma^+\sigma^+$) or ($\sigma^-\sigma^-$) configuration.[1,7,8] The PL peaks exhibit a Zeeman splitting as defined by $E = E_0 \pm \frac{1}{2}g\mu_B B$, where g is the Landé g-factor, $\mu_B$ is the Bohr magneton. The "+" and "-" signs are for the PL peak energies from the K and K' valleys, respectively. The PL peaks in the magneto-PL spectra of the ($\sigma^-\sigma^-$) (Fig. 4a) configuration exhibits a linear blue shift due to the valley Zeeman effects,[20,44-53] and the size of the shift in a fixed B field is determined by the g-factor. It is evident from Fig. 4a that the $X_i$ peak has the largest g-factor because the shift of the $X_i$ peak has the steepest slope as a function of the magnetic field. Quantitatively, we can extract the g-factor from the experimental data of ($\sigma^-\sigma^-$) and ($\sigma^+\sigma^+$) by calculating the Zeeman shift difference between the K and K' valleys (see Supplementary Note 1 in SI), and the results are shown in Fig. 4b. The experimentally extracted g-factor for $X_i$ is ~ -12.5, significantly larger than that of the bright exciton (~ -3.6), trions (~ -4.5 and ~ -4.0 for intervalley trion and intravalley trions, respectively), and the dark exciton (~ -9.3).[19,27,32,33,45] Theoretically, the g-factor can be determined by the overall contribution of the spin, valley, and orbital momentum components combined, and the theoretically expected value can be used to illustrate the nature of the quasiparticle. Considering the charge-neutral nature of the quasiparticle associated with PL peak $X_i$ and the large g factor, the only possibility for the quasiparticle is the intervalley exciton, as schematically shown in Fig. 3c. Based on a non-interacting picture (see SI), the intervalley exciton is supposed to have a g-factor of -12, much larger in magnitude than the theoretically expected value for the bright exciton (g-factor of -4) and the dark exciton (g-factor of -8). Our experimentally extracted g-factor for the $X_i$ peak is in excellent agreement with the theoretically expected value of the intervalley exciton. In contrast, we expect significantly different g-factor for the Q-K exciton, as the orbital component of the conduction band at Q point is different from that of the K valley, and the Berry phase of the Q point is zero[54].

Based on the results from the gate and magnetic-field dependent PL spectra, we identify the $X_i$ peak to be associated with the intervalley exciton as shown in Fig. 3c. The intervalley exciton consists of an electron and a hole residing in different valleys, and the direct recombination is forbidden due to their momentum mismatch. Therefore, additional quasi-particles with momentum **K** must be involved in the emission process to satisfy momentum conservation. Interestingly, the $X_i$ peak exhibits a valley polarization as large as ~ 36%, as observed in valley-resolved PL spectroscopy (Fig. 5a). The valley polarization of the $X_i$ requires the quasi-particle to comply with three-fold rotation symmetry and have well-defined PAM. Furthermore, the energy of peak $X_i$ is highly reproducible in all four samples (with respect to the bright exciton). We thus conclude that the involved quasi-particle should be intrinsic to monolayer $WSe_2$, and it is likely to be a phonon with momentum **K**. Such phonon-assisted emission of the intervalley exciton is already illustrated in Fig. 1, where the intervalley exciton scatters to a virtual intravalley state through emitting a phonon.

To gain further knowledge of this phonon mode, we analyze the behavior of $X_i$ peak in detail. The intervalley dark exciton is theoretically expected to have an energy about 6 meV higher than that of the intravalley dark exciton due to the exchange interaction,[4,20,35,37,45,55] and our measurement of the splitting between the intervalley and intravlley n-trions reveals a value of ~ 7 meV (energy difference between $X_1^-$ and $X_2^-$). Combined with the experimentally determined separation of 16 meV between the intravalley dark exciton ($X_D$) and the $X_i$ peak, we determine the energy of the phonon mode to be around 23 meV. It is worth noting that, the temperature dependent PL spectra (Fig. S1e in SI) show that, despite the shift of the $X_D$ and $X_i$ as a function of the temperature, the energy separation between these two peaks remains a constant, ~ 16 meV, consistent with the phonon assisted recombination picture (assuming that the exchange interaction is not a sensitive function of temperature).

In addition, the peculiar valley polarization of the $X_i$ peak bears important information about the symmetry properties of the phonon mode. Because holes have much longer valley lifetime than electrons in $WSe_2$,[23] the bright excitons in the K valley created by $\sigma^+$ excitation will mainly generate intervalley excitons with holes maintained in the K valley and electrons scattered into the K' valley. As a result, if the phonon mode does not have PAM, the intervalley dark exciton emission is expected to show the opposite circular helicity as the excitation light,[21] similar to the case of interlayer exciton emission in AB-stacked $WSe_2/MoSe_2$ heterostructure.[56] In contrast, a large positive circular helicity is observed experimentally for the $X_i$ peak, suggesting that the phonon mode with momentum **K** should have additional PAM of -2 (or equivalently, +1).

Theoretical analysis[24] reveals that at the corner of Brillouin Zone (K or K' point), the threefold rotational symmetry endows phonon eigenmodes with a PAM which includes both orbital and spin parts. The orbital PAM for sublattice W and Se at K point can be determined through phase change under counterclockwise 120° rotations, and the total PAM of each phonon mode can be obtained after further analyzing the spin PAM of each phonon eigenmode (see Fig. S6 and Table S3 in SI). Since the system is symmetric with regard to a mirror operation with respect to the monolayer $WSe_2$ plane (M=1), we determine the phonon mode involved in the intervalley exciton recombination process to be the chiral phonon mode LO(E') at K, with the energy calculated to be 24.6 meV (Table S3 in SI). This chiral phonon mode features a unidirectional circular rotation as illustrated in the inset of Fig.1, and it has been previously observed experimentally by investigating the transition between A and B exciton states with transient infrared spectroscopy.[25] The observation of intervalley exciton PL here provides another direct evidence of the chiral phonon and its intriguing capability in manipulating the symmetry and valley selection rules of exciton states. In addition, owing to the simplicity of the intervalley exciton emission process without involving an additional infrared photon, the energy of the chiral phonon mode (~ 23 meV) can be accurately determined, which is in excellent agreement with the theoretical prediction of 24.6 meV. It is worth noting that the dark exciton emission in Fig. 5a is not valley polarized, consistent with our expectation, since the observed $X_D$ emission arises from out-of-plane dipole radiation, which is *p*-polarized and has equal intensity from the left-polarized and right-polarized PL in our detection scheme.[27]

Finally, we performed time-resolved PL measurements using the time-correlated single photon counting (TCSPC) technique. The second harmonic generation (SHG) signal from a Ti:Sapphire oscillator (80 MHz, ~ 120 fs pulse width), centered at 2.756 eV, was used as the excitation source. The direct observation of the sharp intravalley dark exciton PL peak (Fig. 5a) enables time-resolved PL measurements, which reveals the lifetime of the dark exciton. As shown in Fig. 5b, the lifetime of the $X_i$ peak is determined to be around 200 ps. It is significantly longer than that of the bright exciton (~ 5 ps) but comparable to the lifetime of the spin-forbidden intravalley dark exciton (~ 250 ps),[17,19,20,28,36] which rules out the possibility of defects, since they usually possess even longer lifetime.[29,57] The long lifetime of the $X_i$ peak also suggest that the measured high valley polarization (36%, Fig. 5a) arises from decreased channels of valley depolarization, which also rules out the possibility of defects and strongly suggests the angular momentum conservation in the recombination process, consistent with our interpretation involving the chiral phonon.

**CONCLUSIONS**

In summary, we have identified the PL peak associated with the momentum-dark, intervalley exciton through gate and magnetic-field dependent PL spectroscopy at low temperature. The energy and momentum conservation demand a **K** phonon with particular energy to be involved to brighten the intervalley exciton. The robust valley polarization also requires the angular momentum conservation in the phonon assisted recombination process, which can be satisfied with the chiral **K** phonon possessing a well-defined PAM. The long lived intervalley exciton with valley information coded shed light on promising routes of realizing excitonic valleytronics, and the coupling of the intervalley exciton with the chiral phonon could inspire future endeavor of exploiting chiral phonon for valleyspin manipulation.

**METHODS**

**Fabrication of h-BN Encapsulated monolayer $WSe_2$ Device**

We fabricated The h-BN encapsulated monolayer $WSe_2$ device by the well-known dry transfer technique.[26,27] The monolayer $WSe_2$, few-layer graphene and few-layer h-BN were first exfoliated onto 285 nm $SiO_2$/Si substrate, followed by the inspection through optical microscope. Afterward, the PPC (polypropylene carbonate) stamp sequentially picks up the few-layer h-BN, monolayer $WSe_2$, few-layer graphene and another few-layer h-BN. The whole stack of the constructed vdW structure was placed onto pre-patterned Au electrodes, and the PPC is removed by heating up to 90 °C. Then, the whole chip was soaked in chloroform for 2 hours to remove the PPC polymer residue. Finally, we transferred another few-layer graphene flake to work as the top-gate electrode, on top of the top h-BN flake.

**Magneto-PL and Time-Resolved PL Measurements**[26,27]

We applied a confocal micro-PL setup to measure the magneto-PL with the out-of-plane magnetic field. A linear polarized light is converted into circularly polarized light by a quarter waveplate and the laser is focused by a 50X objective (NA: ~ 0.65) to excite the $WSe_2$ sample, with the spot size of ~ 2 μm. The PL is collected with the same objective and goes through the quarter waveplate to be converted into linear light. The assembly of a half waveplate and a linear polarizer is used to distinguish the $\sigma^+\sigma^+$ and $\sigma^-\sigma^-$ configurations and the signal is detected by the CCD camera attached to the spectrometer.

We measured the TRPL using the time-correlated single photon counting (TCSPC) technique, with the pulsed laser excitation centered at 2.756 eV and the excitation power of 50 μW. For each PL peak, we integrated the PL intensity around the center with an uncertainty of $\pm 1$ meV for the time-resolved information. The pulsed laser is the second harmonic generation (SHG) signal from a Ti:Sapphire oscillator (80 MHz, ~ 120 fs pulse width). We performed convolution and extracted the lifetime of different excitonic complexes by a single exponential function $I = Ae^{-t/\tau}$ convolved with the response of the laser as a kernel.

## NOTES

The authors declare no competing financial interests.

## AUTHOR INFORMATION

# These authors contributed equally to this work

**Corresponding Authors:**

*E-mail: shis2@rpi.edu, phyzlf@njnu.edu.cn, thren@sjtu.edu.cn.

## ACKNOWLEDGMENTS

We thank Prof. Feng Wang, Prof. Ji Feng, and Prof. Ronald Hedden for helpful discussions. We acknowledge the support by AFSOR through Grant FA9550-18-1-0312, and we acknowledge Micro and Nanofabrication Clean Room (MNCR) at Rensselaer Polytechnic Institute (RPI) for device fabrication. Z. Li acknowledges supports from the Shanghai Sailing Program (grant no. 19YF1425200) and the National Natural Science Foundation for Young Scientists Fund of China (grant no. 51902196). L.Z. acknowledges support from the National Natural Science Foundation of China (grants no. 11890703, 11574154). S. Tongay acknowledges support from NSF DMR-1552220 and DMR 1838443. C.J. acknowledges support from a Kavli Postdoctoral Fellowship. K.W. and T.T. acknowledge support from the Elemental Strategy Initiative conducted by the MEXT, Japan and the CREST (JPMJCR15F3), JST. Z. Lu. and D.S. acknowledge support from the US Department of Energy (DE-FG02-07ER46451) for magneto-PL work performed at the NHMFL, which is supported by NSF through NSF/DMR-1644779 and the State of Florida. T. Wang and S.-F. Shi acknowledge support from ACS PRF through grant 59957-DNI10. Z. Lian and S.-F. Shi acknowledge support from the New York State Empire State Development's Division of Science, Technology, and Innovation (NYSTAR) through Focus Center-NY-RPI Contract C150117. S.-F. Shi is also supported by a KIP grant from RPI and a VSP grant from NHMFL.

## REFERENCES

(1) Xiao, D.; Liu, G. Bin; Feng, W.; Xu, X.; Yao, W. Coupled Spin and Valley Physics in Monolayers of MoS$_2$ and Other Group-VI Dichalcogenides. *Phys. Rev. Lett.* **2012**, *108*, 196802.

(2) Gong, Z.; Liu, G. Bin; Yu, H.; Xiao, D.; Cui, X.; Xu, X.; Yao, W. Magnetoelectric Effects

and Valley-Controlled Spin Quantum Gates in Transition Metal Dichalcogenide Bilayers. *Nat. Commun.* **2013**, *4*, 2053.

(3) Ye, Z.; Sun, D.; Heinz, T. F. Optical Manipulation of Valley Pseudospin. *Nat. Phys.* **2017**, *13*, 26–29.

(4) Plechinger, G.; Nagler, P.; Arora, A.; Schmidt, R.; Chernikov, A.; Del Águila, A. G.; Christianen, P. C. M.; Bratschitsch, R.; Schüller, C.; Korn, T. Trion Fine Structure and Coupled Spin-Valley Dynamics in Monolayer Tungsten Disulfide. *Nat. Commun.* **2016**, *7*, 12715.

(5) Mak, K. F.; Shan, J. Photonics and Optoelectronics of 2D Semiconductor Transition Metal Dichalcogenides. *Nat. Photonics* **2016**, *10*, 216–226.

(6) Xu, X.; Yao, W.; Xiao, D.; Heinz, T. F. Spin and Pseudospins in Layered Transition Metal Dichalcogenides. *Nat. Phys.* **2014**, *10*, 343–350.

(7) Cao, T.; Wang, G.; Han, W.; Ye, H.; Zhu, C.; Shi, J.; Niu, Q.; Tan, P.; Wang, E.; Liu, B.; Feng, J. Valley-Selective Circular Dichroism of Monolayer Molybdenum Disulphide. *Nat. Commun.* **2012**, *3*, 885–887.

(8) Mak, K. F.; He, K.; Shan, J.; Heinz, T. F. Control of Valley Polarization in Monolayer $MoS_2$ by Optical Helicity. *Nat. Nanotechnol.* **2012**, *7*, 494–498.

(9) Peng, B.; Li, Q.; Liang, X.; Song, P.; Li, J.; He, K.; Fu, D.; Li, Y.; Shen, C.; Wang, H.; Wang, C; Liu, T.; Zhang, L.; Lu, H.; Wang, X.; Zhao, J.; Xie, J.; Wu, M.; Bi, L.; Deng, L.; *et al*. Valley Polarization of Trions and Magnetoresistance in Heterostructures of $MoS_2$ and Yttrium Iron Garnet. *ACS Nano* **2017**, *11*, 12257–12265.

(10) Shin, D.; Hübener, H.; De Giovannini, U.; Jin, H.; Rubio, A.; Park, N. Phonon-Driven Spin-Floquet Magneto-Valleytronics in $MoS_2$. *Nat. Commun.* **2018**, *9*, 638.

(11) Lundeberg, M. B.; Folk, J. A. Harnessing Chirality for Valleytronics. *Science.* **2014**, *346*, 422–423.

(12) Schaibley, J. R.; Yu, H.; Clark, G.; Rivera, P.; Ross, J. S.; Seyler, K. L.; Yao, W.; Xu, X. Valleytronics in 2D Materials. *Nat. Rev. Mater.* **2016**, *1*, 16055.

(13) Zhong, D.; Seyler, K. L.; Linpeng, X.; Cheng, R.; Sivadas, N.; Huang, B.; Schmidgall, E.; Taniguchi, T.; Watanabe, K.; McGuire, M. A.; Yao, W.; Xiao, D.; Fu, K. M. C.; Xu, X. Van Der Waals Engineering of Ferromagnetic Semiconductor Heterostructures for Spin and Valleytronics. *Sci. Adv.* **2017**, *3*, e1603113.

(14) Wang, G.; Chernikov, A.; Glazov, M. M.; Heinz, T. F.; Marie, X.; Amand, T.; Urbaszek, B. Colloquium: Excitons in Atomically Thin Transition Metal Dichalcogenides. *Rev. Mod. Phys.* **2018**, *90*, 21001.

(15) Jin, C.; Ma, E. Y.; Karni, O.; Regan, E. C.; Wang, F.; Heinz, T. F. Ultrafast Dynamics in van Der Waals Heterostructures. *Nat. Nanotechnol.* **2018**, *13*, 994–1003.

(16) Rivera, P.; Seyler, K. L.; Yu, H.; Schaibley, J. R.; Yan, J.; Mandrus, D. G.; Yao, W.; Xu, X. Valley-Polarized Exciton Dynamics in a 2D Semiconductor Heterostructure. *Science.*

2016, *351*, 688–691.

(17) Robert, C.; Lagarde, D.; Cadiz, F.; Wang, G.; Lassagne, B.; Amand, T.; Balocchi, A.; Renucci, P.; Tongay, S.; Urbaszek, B.; Marie, X. Exciton Radiative Lifetime in Transition Metal Dichalcogenide Monolayers. *Phys. Rev. B* **2016**, *93*, 205423.

(18) Lagarde, D.; Bouet, L.; Marie, X.; Zhu, C. R.; Liu, B. L.; Amand, T.; Tan, P. H.; Urbaszek, B. Carrier and Polarization Dynamics in Monolayer $MoS_2$. *Phys. Rev. Lett.* **2014**, *112*, 047401.

(19) Robert, C.; Amand, T.; Cadiz, F.; Lagarde, D.; Courtade, E.; Manca, M.; Taniguchi, T.; Watanabe, K.; Urbaszek, B.; Marie, X. Fine Structure and Lifetime of Dark Excitons in Transition Metal Dichalcogenide Monolayers. *Phys. Rev. B* **2017**, *96*, 155423.

(20) Zhang, X. X.; Cao, T.; Lu, Z.; Lin, Y. C.; Zhang, F.; Wang, Y.; Li, Z.; Hone, J. C.; Robinson, J. A.; Smirnov, D.; Louie, S. G.; Heinz, T. F. Magnetic Brightening and Control of Dark Excitons in Monolayer $WSe_2$. *Nat. Nanotechnol.* **2017**, *12*, 883–888.

(21) Wang, G.; Robert, C.; Glazov, M. M.; Cadiz, F.; Courtade, E.; Amand, T.; Lagarde, D.; Taniguchi, T.; Watanabe, K.; Urbaszek, B.; Marie, X. In-Plane Propagation of Light in Transition Metal Dichalcogenide Monolayers: Optical Selection Rules. *Phys. Rev. Lett.* **2017**, *119*, 047401.

(22) Zhou, Y.; Scuri, G.; Wild, D. S.; High, A. A.; Dibos, A.; Jauregui, L. A.; Shu, C.; De Greve, K.; Pistunova, K.; Joe, A. Y.; Taniguchi, T.; Watanabe, K.; Kim, P.; Lukin, M. D.; Park, H. Probing Dark Excitons in Atomically Thin Semiconductors via Near-Field Coupling to Surface Plasmon Polaritons. *Nat. Nanotechnol.* **2017**, *12*, 856–860.

(23) Kim, J.; Jin, C.; Chen, B.; Cai, H.; Zhao, T.; Lee, P.; Kahn, S.; Watanabe, K.; Taniguchi, T.; Tongay, S.; Crommie, M. F.; Wang, F. Observation of Ultralong Valley Lifetime in $WSe_2$/$MoS_2$ Heterostructures. *Sci. Adv.* **2017**, *3*, e1700518.

(24) Zhang, L.; Niu, Q.; Phonon, V. I. Chiral Phonons at High-Symmetry Points in Monolayer Hexagonal Lattices. *Phys. Rev. Lett.* **2015**, *115*, 115502.

(25) Zhu, H.; Yi, J.; Li, M. Y.; Xiao, J.; Zhang, L.; Yang, C. W.; Kaindl, R. A.; Li, L. J.; Wang, Y.; Zhang, X. Observation of Chiral Phonons. *Science.* **2018**, *359*, 579–582.

(26) Li, Z.; Wang, T.; Jin, C.; Lu, Z.; Lian, Z.; Meng, Y.; Blei, M.; Gao, S.; Taniguchi, T.; Watanabe, K.; Ren, T.; Tongay, S.; Yang, L.; Smirnov, D.; Cao, T.; Shi, S. F. Emerging Photoluminescence from the Dark-Exciton Phonon Replica in Monolayer $WSe_2$. *Nat. Commun.* **2019**, *10*, 2469.

(27) Li, Z.; Wang, T.; Lu, Z.; Jin, C.; Chen, Y.; Meng, Y.; Lian, Z.; Taniguchi, T.; Watanabe, K.; Zhang, S.; Smirnov, D.; Shi, S.-F. Revealing the Biexciton and Trion-Exciton Complexes in BN Encapsulated $WSe_2$. *Nat. Commun.* **2018**, *9*, 3719.

(28) Nagler, P.; Ballottin, M. V.; Mitioglu, A. A.; Durnev, M. V.; Taniguchi, T.; Watanabe, K.; Chernikov, A.; Schüller, C.; Glazov, M. M.; Christianen, P. C. M.; Korn, T. Zeeman Splitting and Inverted Polarization of Biexciton Emission in Monolayer $WS_2$. *Phys. Rev. Lett.* **2018**, *121*, 57402.

(29) You, Y.; Zhang, X. X.; Berkelbach, T. C.; Hybertsen, M. S.; Reichman, D. R.; Heinz, T. F. Observation of Biexcitons in Monolayer WSe$_2$. *Nat. Phys.* **2015**, *11*, 477–481.

(30) Hao, K.; Specht, J. F.; Nagler, P.; Xu, L.; Tran, K.; Singh, A.; Dass, C. K.; Korn, T.; Richter, M.; Knorr, A.; Li, X.; Moody, G. Neutral and Charged Inter-Valley Biexcitons in Monolayer MoSe$_2$. *Nat. Commun.* **2017**, *8*, 15552.

(31) Chen, S.-Y.; Goldstein, T.; Taniguchi, T.; Watanabe, K.; Yan, J. Coulomb-Bound Four- and Five-Particle Intervalley States in an Atomically-Thin Semiconductor. *Nat. Commun.* **2018**, *9*, 3717.

(32) Barbone, M.; Montblanch, A. R.-P.; Kara, D. M.; Palacios-Berraquero, C.; Cadore, A. R.; De Fazio, D.; Pingault, B.; Mostaani, E.; Li, H.; Chen, B.; Watanabe, K.; Taniguchi, T.; Tongay, S.; Wang, G.; Ferrari, A. C.; Atatüre, M. Charge-Tuneable Biexciton Complexes in Monolayer WSe$_2$. *Nat. Commun.* **2018**, *9*, 3721.

(33) Ye, Z.; Waldecker, L.; Ma, E. Y.; Rhodes, D.; Antony, A.; Kim, B.; Zhang, X. X.; Deng, M.; Jiang, Y.; Lu, Z.; Smirnov, D.; Watanabe, K.; Taniguchi, T.; Hone, J.; Heinz, T. F. Efficient Generation of Neutral and Charged Biexcitons in Encapsulated WSe$_2$ Monolayers. *Nat. Commun.* **2018**, *9*, 3718.

(34) Kim, M. S.; Yun, S. J.; Lee, Y.; Seo, C.; Han, G. H.; Kim, K. K.; Lee, Y. H.; Kim, J. Biexciton Emission from Edges and Grain Boundaries of Triangular WS$_2$ Monolayers. *ACS Nano* **2016**, *10*, 2399–2405.

(35) Jones, A. M.; Yu, H.; Schaibley, J. R.; Yan, J.; Mandrus, D. G.; Taniguchi, T.; Watanabe, K.; Dery, H.; Yao, W.; Xu, X. Excitonic Luminescence Upconversion in a Two-Dimensional Semiconductor. *Nat. Phys.* **2016**, *12*, 323–327.

(36) Singh, A.; Tran, K.; Kolarczik, M.; Seifert, J.; Wang, Y.; Hao, K.; Pleskot, D.; Gabor, N. M.; Helmrich, S.; Owschimikow, N.; Woggon, U.; Li, X. Long-Lived Valley Polarization of Intravalley Trions in Monolayer WSe$_2$. *Phys. Rev. Lett.* **2016**, *117*, 257402.

(37) Yu, H.; Liu, G. Bin; Gong, P.; Xu, X.; Yao, W. Dirac Cones and Dirac Saddle Points of Bright Excitons in Monolayer Transition Metal Dichalcogenides. *Nat. Commun.* **2014**, *5*, 3876.

(38) Shang, J.; Shen, X.; Cong, C.; Peimyoo, N.; Cao, B.; Eginligil, M. Observation of Excitonic Fine Structure in a 2D Transition-Metal Dichalcogenide Semiconductor. *ACS Nano* **2015**, *9*, 647–655.

(39) Park, K. D.; Jiang, T.; Clark, G.; Xu, X.; Raschke, M. B. Radiative Control of Dark Excitons at Room Temperature by Nano-Optical Antenna-Tip Purcell Effect. *Nat. Nanotechnol.* **2018**, *13*, 59–64.

(40) Zhang, X. X.; You, Y.; Zhao, S. Y. F.; Heinz, T. F. Experimental Evidence for Dark Excitons in Monolayer WSe$_2$. *Phys. Rev. Lett.* **2015**, *115*, 257403.

(41) Hsu, W. T.; Lu, L. S.; Wang, D.; Huang, J. K.; Li, M. Y.; Chang, T. R.; Chou, Y. C.; Juang, Z. Y.; Jeng, H. T.; Li, L. J.; Chang, W. H. Evidence of Indirect Gap in Monolayer WSe$_2$. *Nat. Commun.* **2017**, *8*, 2.

(42) Liu, E.; van Baren, J.; Lu, Z.; Altaiary, M. M.; Taniguchi, T.; Watanabe, K.; Smirnov, D.; Lui, C. H. Gate Tunable Dark Trions in Monolayer $WSe_2$. *Phys. Rev. Lett.* **2019**, *123*, 027401.

(43) Tang, Y.; Mak, K. F.; Shan, J. Long Valley Lifetime of Dark Excitons in Single-Layer $WSe_2$. *Nat. Commun.* **2019**, *10*, 4047.

(44) Macneill, D.; Heikes, C.; Mak, K. F.; Anderson, Z.; Kormányos, A.; Zólyomi, V.; Park, J.; Ralph, D. C. Breaking of Valley Degeneracy by Magnetic Field in Monolayer $MoSe_2$. *Phys. Rev. Lett.* **2015**, *114*, 037401.

(45) Aivazian, G.; Gong, Z.; Jones, A. M.; Chu, R.-L.; Yan, J.; Mandrus, D. G.; Zhang, C.; Cobden, D.; Yao, W.; Xu, X. Magnetic Control of Valley Pseudospin in Monolayer $WSe_2$. *Nat. Phys.* **2015**, *11*, 148–152.

(46) Srivastava, A.; Sidler, M.; Allain, A. V.; Lembke, D. S.; Kis, A.; Imamoglu, A. Valley Zeeman Effect in Elementary Optical Excitations of Monolayer $WSe_2$. *Nat. Phys.* **2015**, *11*, 141–147.

(47) Li, Y.; Ludwig, J.; Low, T.; Chernikov, A.; Cui, X.; Arefe, G.; Kim, Y. D.; Van Der Zande, A. M.; Rigosi, A.; Hill, H. M.; Kim, S. H.; Hone, J.; Li, Z.; Smirnov, D.; Heinz, T. F. Valley Splitting and Polarization by the Zeeman Effect in Monolayer $MoSe_2$. *Phys. Rev. Lett.* **2014**, *113*, 266804.

(48) Yuan, H.; Bahramy, M. S.; Morimoto, K.; Wu, S.; Nomura, K.; Yang, B.-J.; Shimotani, H.; Suzuki, R.; Toh, M.; Kloc, C.; Xu, X.; Arita, R.; Nagaosa, N.; Iwasa, Y. Zeeman-Type Spin Splitting Controlled by an Electric Field. *Nat. Phys.* **2013**, *9*, 563–569.

(49) Rybkovskiy, D. V.; Gerber, I. C.; Durnev, M. V. Atomically Inspired k · p Approach and Valley Zeeman Effect in Transition Metal Dichalcogenide Monolayers. *Phys. Rev. B* **2017**, *95*, 155406.

(50) Stier, A. V.; McCreary, K. M.; Jonker, B. T.; Kono, J.; Crooker, S. A. Exciton Diamagnetic Shifts and Valley Zeeman Effects in Monolayer $WS_2$ and $MoS_2$ to 65 Tesla. *Nat. Commun.* **2016**, *7*, 10643.

(51) Jiang, C.; Liu, F.; Cuadra, J.; Huang, Z.; Li, K.; Rasmita, A.; Srivastava, A.; Liu, Z.; Gao, W. B. Zeeman Splitting via Spin-Valley-Layer Coupling in Bilayer $MoTe_2$. *Nat. Commun.* **2017**, *8*, 802.

(52) Nagler, P.; Ballottin, M. V.; Mitioglu, A. A.; Mooshammer, F.; Paradiso, N.; Strunk, C.; Huber, R.; Chernikov, A.; Christianen, P. C. M.; Schüller, C.; Korn, T. Giant Zeeman Splitting Inducing Near-Unity Valley Polarization in van Der Waals Heterostructures. *Nat. Commun.* **2017**, *8*, 1551.

(53) Arora, A.; Schmidt, R.; Schneider, R.; Molas, M. R.; Breslavetz, I.; Potemski, M.; Bratschitsch, R. Valley Zeeman Splitting and Valley Polarization of Neutral and Charged Excitons in Monolayer $MoTe_2$ at High Magnetic Fields. *Nano Lett.* **2016**, *16*, 3624–3629.

(54) Liu, G.-B.; Xiao, D.; Yao, Y.; Xu, X.; Yao, W. Electronic Structures and Theoretical Modelling of Two-Dimensional Group-VIB Transition Metal Dichalcogenides. *Chem. Soc. Rev.* **2015**, *44*, 2643–2663.

(55) Steinhoff, A.; Florian, M.; Singh, A.; Tran, K.; Kolarczik, M.; Helmrich, S.; Achtstein, A. W.; Woggon, U.; Owschimikow, N.; Jahnke, F.; Li, X. Biexciton Fine Structure in Monolayer Transition Metal Dichalcogenides. *Nat. Phys.* **2018**, *14*, 1199–1204.

(56) Hsu, W. T.; Lu, L. S.; Wu, P. H.; Lee, M. H.; Chen, P. J.; Wu, P. Y.; Chou, Y. C.; Jeng, H. T.; Li, L. J.; Chu, M. W.; Chang, W. H. Negative Circular Polarization Emissions from WSe$_2$/MoSe$_2$ Commensurate Heterobilayers. *Nat. Commun.* **2018**, *9*, 1356.

(57) Moody, G.; Tran, K.; Lu, X.; Autry, T.; Fraser, J. M.; Mirin, R. P.; Yang, L.; Li, X.; Silverman, K. L. Microsecond Valley Lifetime of Defect-Bound Excitons in Monolayer WSe$_2$. *Phys. Rev. Lett.* **2018**, *121*, 057403.

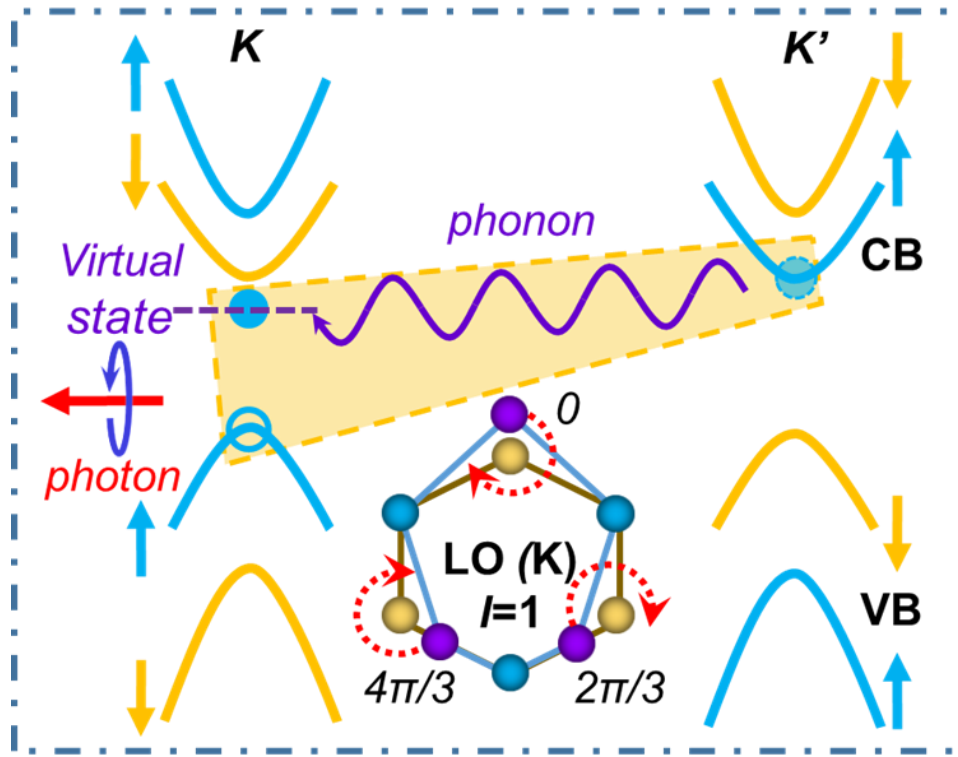

**Figure 1. Scheme of the intervalley exciton and chiral phonon coupling.** Assuming the initial population of excitons in the K valley, the intervalley exciton consists of one electron in the K' valley and one hole in the K valley. The electron transitions to a virtual state in the K valley by emitting a chiral phonon and then recombines with the hole in the valence band of the same valley, emitting a photon with certain helicity. Inset: schematic representation of the chiral phonon mode. Blue sphere: W atom. Yellow and purple spheres are Se atoms in the equilibrium state and the vibration state, respectively.

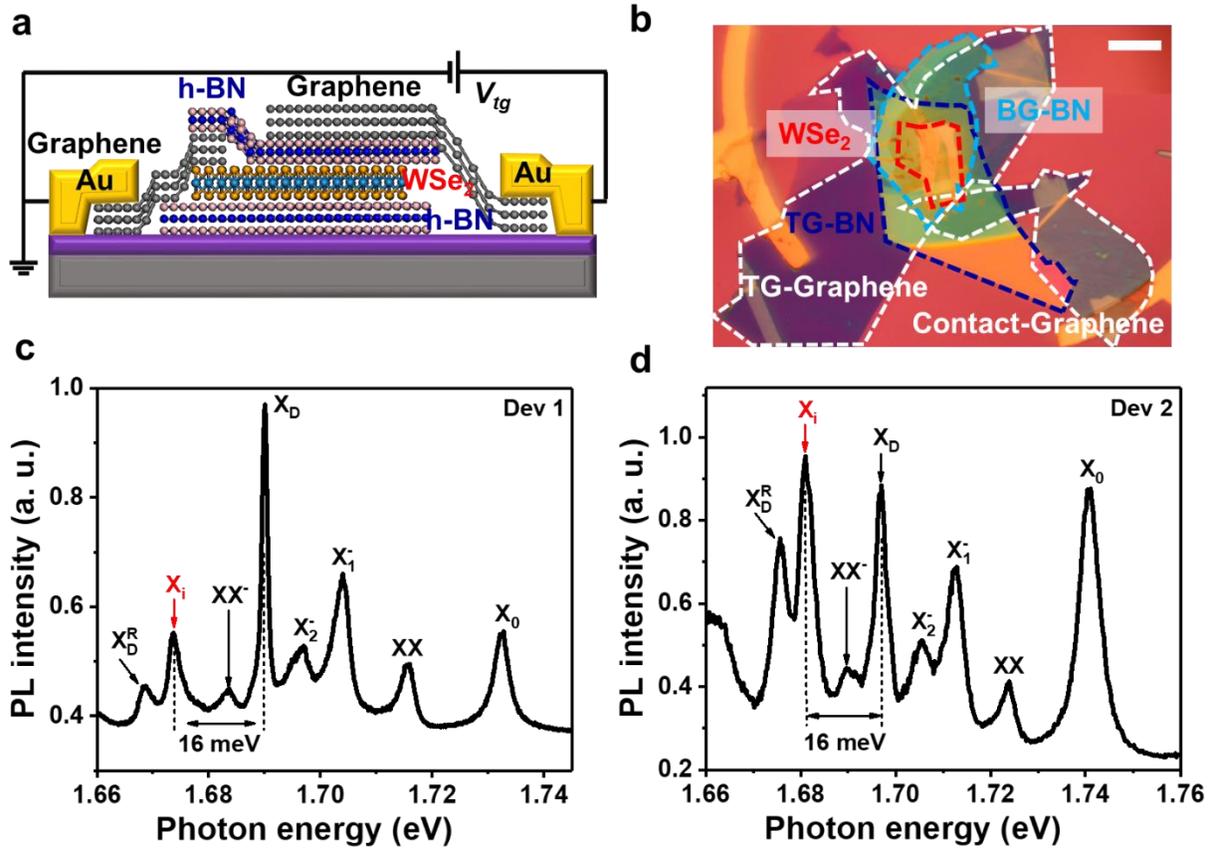

**Figure 2. Low-temperature PL spectra of BN encapsulated monolayer WSe$_2$.** (a) Schematic of the BN encapsulated monolayer WSe$_2$ device, with the carrier density controlled by a top gate. (b) Optical microscope image of a fabricated device, scale bar: 20 µm. (c) PL spectra of device1 at 4.2 K with the gate voltage of -0.44 V. A CW laser centered at 1.959 eV was applied as the excitation source, with the excitation power of 40 µW. (d) PL spectra of device 2 at 4.2 K with no gate voltage applied. CW lasers centered at 1.959 eV for (c) and 1.879 eV for (d) were applied as the excitation source, with excitation power of 40 µW and 60 µW for (c) and (d), respectively.

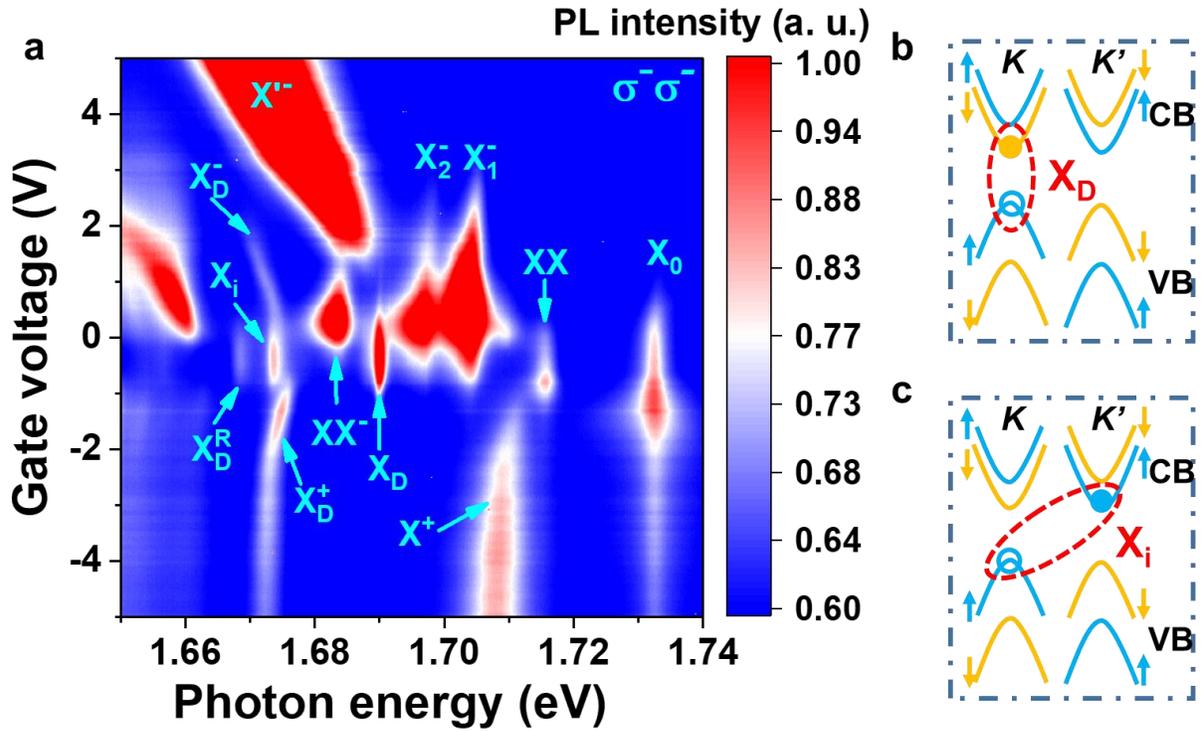

**Figure 3. Gate-voltage dependent PL spectra at low temperature.** (a) Color plot of PL spectra of device 1 at 4.2 K as a function of the top gate voltage. The excitonic complexes are all well-resolved, including the bright exciton ($X_0$), the biexciton (XX) and charge biexciton ($XX^-$), the two negative trions ($X_1^-$ and $X_2^-$), the dark exciton ($X_D$) and the dark trions ($X_D^+$ and $X_D^-$), and the dark exciton phonon replica ($X_D^R$). The color represents the PL intensity. The spectra were obtained by photoexcitation with a CW laser centered at 1.879 eV, with the excitation power of 100 µW. (b-c) Schematic representations of intravalley spin-forbidden dark exciton ($X_D$) and intervalley exciton ($X_i$).

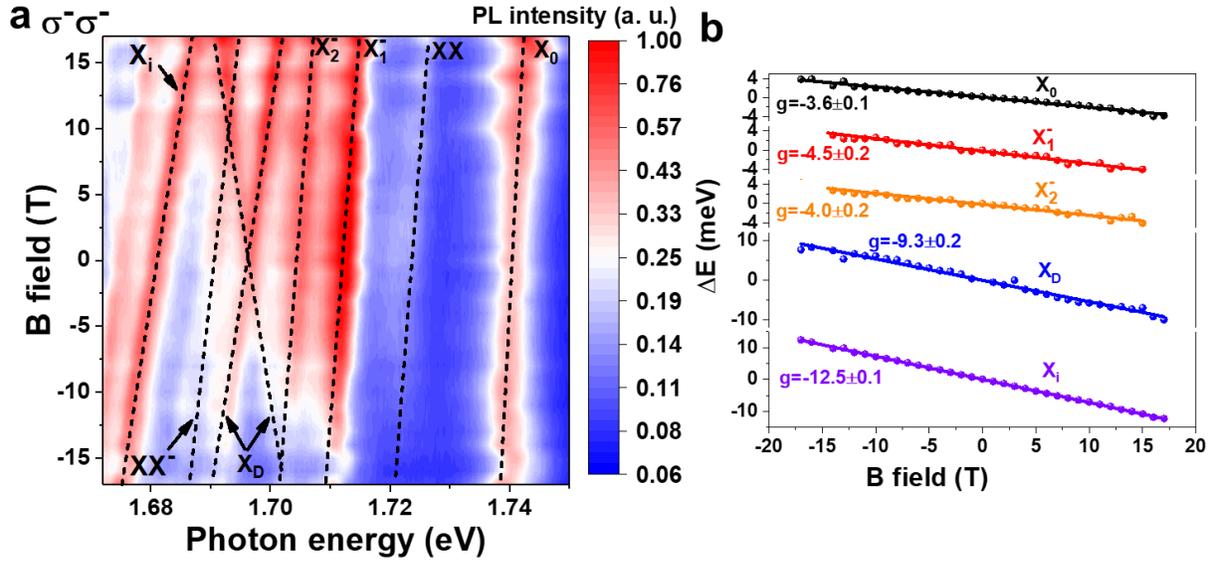

**Figure 4. Magnetic-field dependent PL spectra at low temperature.** (a) Color plot of the PL spectra of device 2 at 4.2 K as a function of the out-of-plane magnetic field for (σ⁻σ⁻) configuration. The spectra were obtained with photoexcitation of a CW laser centered at 1.959 eV, with excitation power of 40 µW. (b) Extracted g-factor from a linear fit of the Zeeman splitting obtained from σ⁻σ⁻ and σ⁺σ⁺ configurations

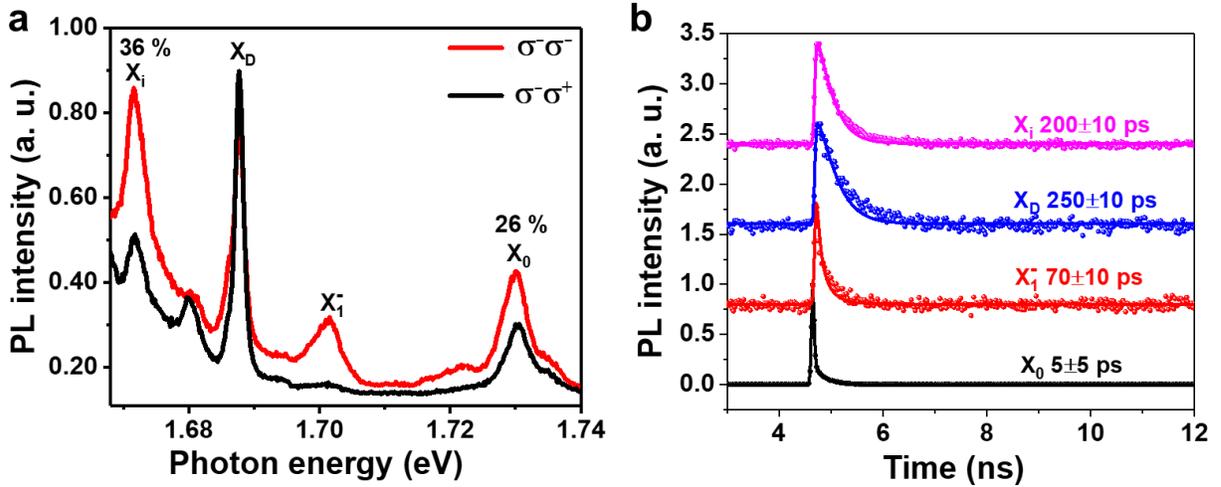

**Figure 5. Valley polarization and time-resolved PL.** (a) Valley-resolved PL spectra at 43 K with left circularly polarized excitation (σ⁻) and left (σ⁻) or right (σ⁺) circularly polarized detection. The spectra were obtained with the photoexcitation of a CW laser centered at 1.797 eV, with the excitation power of 100 µW. The valley polarization, defined as $P = \frac{I_{\sigma^-} - I_{\sigma^+}}{I_{\sigma^-} + I_{\sigma^+}}$, where $I$ stands for PL intensity, is ~ 26% for $X_0$ and ~ 36 % for $X_i$. (b) Time-resolved PL spectra of the distinct PL peaks shown in (a). The spectra were obtained with the pulsed laser excitation (pulse width ~ 120 fs) centered at 2.756 eV, with the excitation power of 50 µW. The lifetime of each excitonic complex is obtained through the convolution with the response from the laser pulse (see SI).

**TOC**

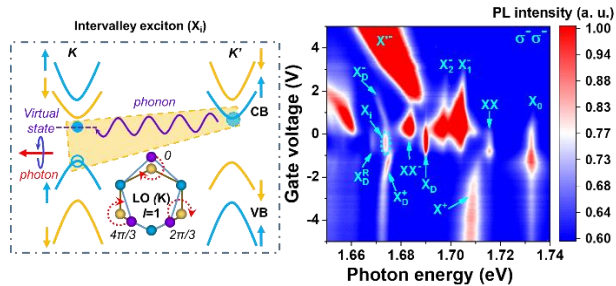